\documentclass[apjl]{emulateapj}

\usepackage{graphicx}
\usepackage{ulem}
\usepackage{url}
\usepackage{threeparttable}
\usepackage{xcolor}
\usepackage{soul}
\usepackage{booktabs}
\usepackage[]{natbib}
\usepackage{amsmath}
\usepackage[colorlinks, linkcolor=black, anchorcolor=green, citecolor=blue]{hyperref}

\usepackage{txfonts}
\usepackage{adjustbox}

\slugcomment{}

\shorttitle{Outflow detection in a 70 $\mu$m dark high-mass core}
\shortauthors{Feng et al.}

\begin{document}


\title{ Outflow detection in a 70 $\mu$m dark high-mass core}



\author{Siyi Feng (Chinese Name)\altaffilmark{1,2}, Henrik Beuther\altaffilmark{2}, Qizhou Zhang (Chinese Name)\altaffilmark{3}, Hauyu Baobab Liu  (Chinese Name)\altaffilmark{4}, Zhiyu Zhang (Chinese Name)\altaffilmark{4, 5},   Ke
Wang (Chinese Name)\altaffilmark{4},   \and Keping Qiu  (Chinese Name)\altaffilmark{6,7} }

\affil{$^1$ Max-Planck-Institut f\"ur Extraterrestrische Physik, 
Gie{\ss}enbachstra{\ss}e 1,  D-85748,  Garching bei M\"unchen, Germany}
\email{syfeng@mpe.mpg.de}

\affil{$^2$ Max-Planck-Institut f\"ur Astronomie, K\"onigstuhl 17,  D-69117,  Heidelberg, Germany}

\affil{$^3$ Harvard-Smithsonian Center for Astrophysics, 60 Garden Street, Cambridge MA 02138, USA} 
\affil{$^4$ European Southern Observatory, Karl-Schwarzschild-Str. 2,  D-85748,  Garching bei M\"unchen, Germany} 
\affil{$^5$  Institute for Astronomy, University of Edinburgh, Royal Observatory, Blackford Hill, Edinburgh EH9 3HJ, UK}

\affil{$^6$ School of Astronomy and Space Science, Nanjing University, 22 Hankou Road, Nanjing 210093, China}
\affil{$^7$ Key Laboratory of Modern Astronomy and Astrophysics (Nanjing University), Ministry of Education, Nanjing 210093, China}


\begin{abstract} 

We present observations towards  a high-mass ($\rm >40\,M_{\odot}$), low luminosity ($\rm <10\,L_{\odot}$) $\rm 70\,\mu$m dark molecular core G\,28.34\,S-A at 3.4\,mm, using the IRAM 30\,m telescope and the NOEMA interferometer.
We report the detection of $\rm SiO$ $J=\rm 2\rightarrow1$ line emission, which is  spatially resolved in this source at a linear resolution of $\sim$0.1\,pc, while the 3.4\,mm  continuum image  does not resolve any internal sub-structures. The SiO emission exhibits two W-E oriented lobes centring on the continuum peak. Corresponding to the red-shifted and blue-shifted gas with velocities up to $\rm 40\,km\,s^{-1}$ relative to the quiescent cloud, these lobes clearly indicate the presence of a strong bipolar outflow from this $\rm 70\,\mu$m dark core,  a source previously considered as one of the best candidates of ``starless" core. 
Our SiO detection is consistent with ALMA archival data of $\rm SiO$ $J=\rm 5\rightarrow4$, whose high-velocity blue-shifted gas reveals a more compact lobe spatially closer to the dust center. 
This outflow indicates that the central source may be  in an early evolutionary stage of forming a high-mass protostar. 
We also find that the low-velocity components (in the range of $\rm V_{lsr}$$\rm _{-5}^{+3}\,km\,s^{-1}$) have an extended, NW-SE oriented distribution. 
Discussing the possible accretion scenarios of the outflow-powering young stellar object, we argue that the molecular line emission and the molecular outflows may provide a better indication of the accretion history when forming young stellar object, than that from a snapshot observations of the present bolometric luminosity. This is particularly significant for the cases of episodic accretion, which may occur during the collapse of the parent molecular core. 


\end{abstract}


\section{INTRODUCTION}\label{intro}
Whether high-mass stars form via a quick global collapse of dense molecular gas core, followed by the formation of a stellar object \citep{tan02,mckee03,krumholz07,gong11,kuiper13}, or whether they form via  competitive accretion of a cluster of low-mass (proto)stars \citep{bonnell01,bonnell04,bonnell07}, remains a fundamental question in the field of high-mass cluster-formation \citep{tan14}.
Identifying and then resolving the initial conditions of high-mass star-formation is the first step towards addressing this question { (e,g., \citealt{zhang09}).}
Candidates of initial high-mass star-forming regions are expected to be embedded in the dense  ($\rm n\ge10^3\text{--}10^5~cm^{-3}$, \citealt{teyssier02,rathborne06,butler09,vasyunin09,ragan09,wangk11}), cold ($\rm T<20\,K$, \citealt{carey98,sridharan05,pillai06,wangy08,wangk12,wienen12,chira13}), and low luminosity molecular clouds,  which have a high dust extinction and a short free-fall collapse timescales  (on the order of $\rm 5\times10^4$\,yrs, e.g., \citealt{russeil10,tackenberg12}).
The so called infrared dark molecular clouds (IRDC) unveiled by the previous {MSX} (e.g.,
 \citealp{egan98,simon06}) and {\it Spitzer} near-mid infrared survey (e.g., GLIMPSE, \citealp{robitaille08,cyganowski08}; MIPSGAL, \citealp{carey09}), have provided excellent road maps to look for these objects.

\begin{figure*}
\centering

\includegraphics[width=20cm]{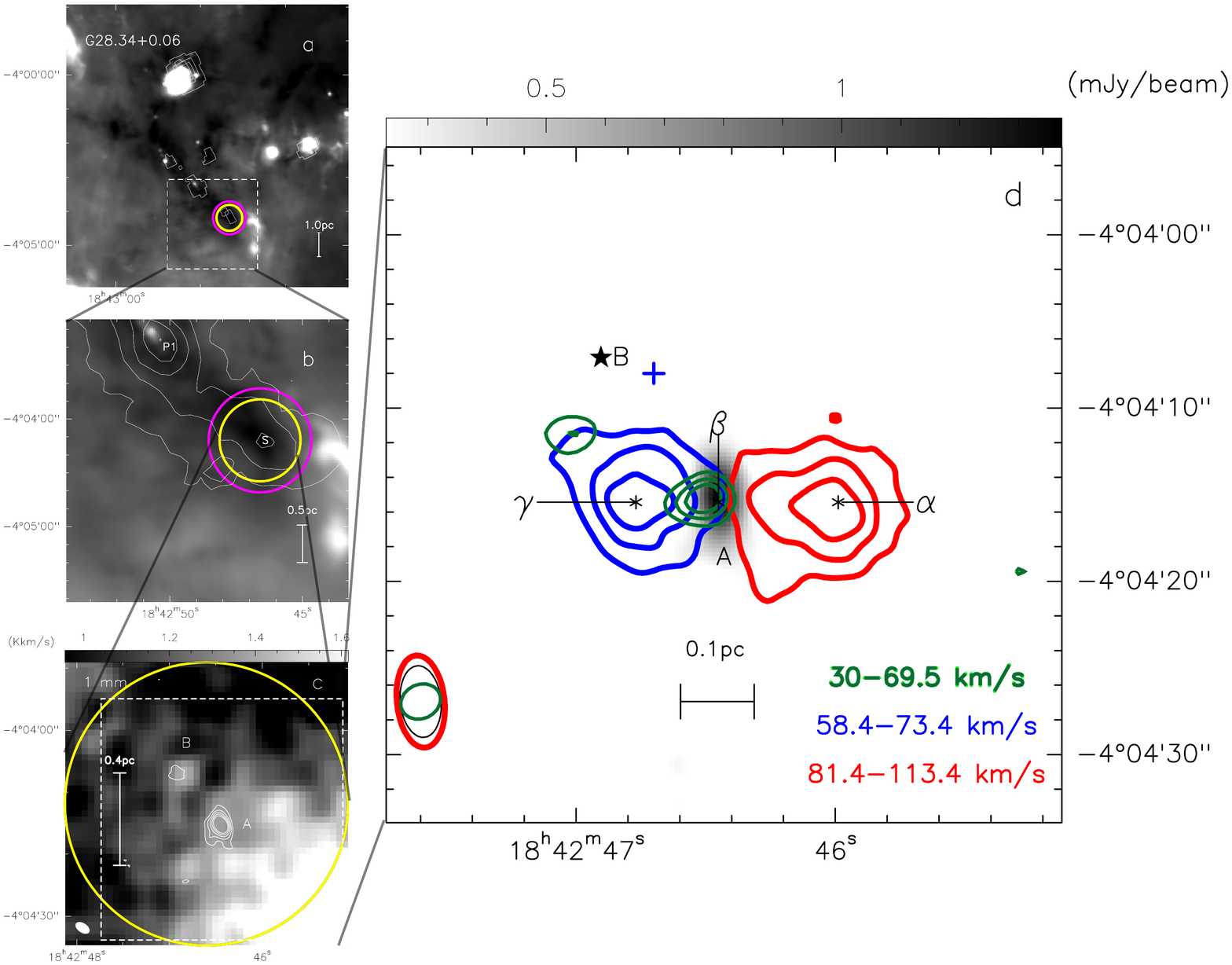}
\caption[]{Compilation of the observations on G\,28.34\,S. 
a) and b). {Greyscale} map of the {dust emission} observed by {\it Herschel}  at $70\,\mu$m \citep{ragan12}.  The white contours  show continuum emission observed by ATLASGAL at $870\,\mu$m  \citep{schuller09}. {White contours in a) start from 15$\sigma$ rms  and continue in 15$\sigma$ rms steps ($\rm \sigma=50\,mJy/beam$), in b) start from 5$\sigma$ rms  and continue in 5$\sigma$ rms steps.  We mark two continuum peaks: P1 \citep{wangk11,wangk12} and S (our targeted source)}. 
c). White contours show continuum observed by SMA COMP configuration at 1.1\,mm, overlaying  the moment 0 greyscale map  of highly depleted ($\sim$15) $\rm C^{18}O$ $J=\rm 2\rightarrow1$ (integrated through its FWZI)  from IRAM 30\,m observations.  Contours start from 5$\sigma$ rms and continue in 5$\sigma$ rms steps ($\rm \sigma=1\,mJy/beam$). 
d). Greyscale map of the continuum observed by NOEMA at 3.4\,mm. Red and blue contours show the red-/blue-shifted gas of $\rm SiO$ $J=\rm 2\rightarrow1$ ($\rm V_{lsr}=78.4\,km\,s^{-1}$) obtained from the combination of NOEMA and 30\,m observations. {The integration range cover the entire red-/blue- shifted line wing.} 
The green contours show the blue-shifted gas of $\rm SiO$ $J=\rm 5\rightarrow4$ from ALMA Cycle 0 public data (the velocity range is only covered { up to  $\rm 69.5\,km\,s^{-1}$} by observations). 
{$\alpha$, $\beta$, and $\gamma$ mark the emission peak of the red-shifted lobe, the  continuum peak, and the emission peak of the blue-shifted lobe, respectively. } 
All the colored contours in figure d start from 4$\sigma$ rms and continue in 4$\sigma$ rms steps  ($\rm \sigma= 88\,mJy/beam$  (red), $\rm 49\,mJy/beam$ (blue), $\rm 35\,mJy/beam$ (green)). The blue cross denotes the $\rm H_2O$ maser at $\rm 59.5\,km\,s^{-1}$ \citep{wangy06}.
In figures b and c, the SMA/NOEMA/NOEMA+30m synthesized beam is in the bottom left.  The yellow circle show the primary beam of SMA at 1.1\,mm, the magenta circle shows the primary beam of NOEMA at 3.4\,mm. 

\label{source}}
\end{figure*}

The $\rm70\,\mu$m dark molecular clump G28.34\,S  \citep{ragan12,ragan13} was discovered around the southern edge of the filamentary IRDC G28.34+0.06 { ($\sim$4.7\,kpc, \citealt{carey98,carey00},  Figure~\ref{source}a and \ref{source}b)}.
Previous studies revealed that this source has high molecular gas density ($\rm >10^5cm^{-3}$, \citealp{butler09}), {low luminosity with a starless-core like SED (10 $\rm L_{\odot}$, \citealp{ragan13}),} low temperature (13--16\,K from SABOCA 350 survey and SPIRE 500), and high deuteration \citep{chenhr10,fontani11}, indicating its  early evolutionary stage.
Our previous observations \citep{feng16} towards this region further reported a high ionization ratio ($\rm >10^{-7}$), a large number of nitrogen bearing species, and significant CO depletion ($\sim15$, Figure~\ref{source}c).  The latter strongly supported by an anti-correlated distribution between  CO and $\rm N_2H^+/NH_2D/H^{13}CO^+$ on scales of 0.8\,pc.
In fact, this region has been considered as one of the best candidates  hosting two high-mass starless cores (e.g. \citealp{chenhr10,tan13,kong16}). 
{However, we surprisingly found $\rm > 5\sigma$ emission of $\rm SiO~J = 2 \rightarrow1$ with $>$0.2\,pc extent  in a 30\,m line survey \citep{feng16}. This line emission, coincident with a {varying} $\rm H_2O$ maser 
 \citep{wangy06,wangk12}, argued that star-formation may have already begun.}

{Outflows are usually considered  important  to distinguish between prestellar and protostellar objects. 
While low-velocity SiO may be released to the gas-phase by large-scale mechanisms such as cloud-cloud collision \citep{jimenez09,jimenez10},
high-velocity SiO  (with broad line wings $\rm >20\,km\,s^{-1}$) regularly observed in high-mass protostars are commonly used to reveal outflows (e.g., \citealt{schilke97,beuther04,qiu07,liu12,duarte14,lopez16}).  Therefore, the line profile of SiO on the core-scale (0.1\,pc) is crucial to judge whether protostellar object(s) are already  in the $\rm 70\,\mu m$ dark G\,28.34\,S.
}



Here we present direct evidence for a spatially resolved molecular outflow in the high-mass molecular gas core G28.34\,S-A (hereafter S-A\footnote{The entire G28.34\,S region corresponds to the ``leaf" No. 5 from dendrogram of SABOCA $\rm 350\,\mu m$ in \citet{ragan13}. S-A and S-B in this paper correspond to C1-S and C1-N in \citet{tan13}, S-A corresponds to MM9 in \citet{rathborne06,chenhr10}.}, Figure~\ref{source}c), using the combined NOEMA (NOrthern Extended Millimeter Array) and IRAM 30\,m telescope observations of the $\rm SiO$ $J=\rm 2\rightarrow1$ line.


\section{Observations and archival data}\label{sec:observation}

\subsection{NOEMA}
To explore the spatial origin of this {strongly emitting} $\rm SiO$ $J=\rm 2\rightarrow1$ line,  we observed G28.34\,S using NOEMA at 87.7\,GHz (3.4\,mm) in its B (with 6-7 working antennae on different days), C  (5 antennae), and D  (5-6 antennae) configurations during March 21 to May 23, 2015. With projected baselines of 24--404 m, the observations are sensitive to structures up to $18''$, and the primary beam is 58\arcsec~ at 3.4\,mm.
For all  observations a common phase center was used,  {at $\rm 18^h42^m46^s.597$, $\rm -04^{\circ}04^{'}11^{''}.940$ (J2000).} The precipitable water vapor  (PWV) varied between 2 and 6 mm during the observations.
Standard interferometric calibrations were performed during the observations, using quasars 1741-038 and 1827+062 as gain calibrators, 3C\,273 or 0923+392 as the bandpass calibrator, and MWC\,349, 1749+096 or 3C\,273 as the flux calibrator. The uncertainty of absolute flux scale is estimated to be correct to within $\rm \sim 5\%$.

$\rm SiO$ $J=\rm 2\rightarrow1$ was covered by a narrow-band correlator unit with a channel width of 0.078 MHz ($\rm 0.275\,km\,s^{-1}$). Continuum emission was {observed} by the wide-band receiver (WIDEX),
covering 85.9--89.5\,GHz with a channel width of 1.95 MHz. 
Data calibration is performed using the GILDAS\footnote{http://www.iram.fr/IRAMFR/GILDAS} package.
The 3.4 mm continuum image is made by averaging the line-free channels of the WIDEX band. We used the natural  weighting to achieve a better signal-to-noise ratio. The final 3.4 mm continuum image has a synthesised beam of $\rm 4.15\arcsec\times2.35\arcsec$ (PA=-171$\rm ^\circ$) and
the $\rm 1\sigma$ rms value is  $\rm 0.08\,mJy beam^{-1}$.

\subsection{IRAM 30~m telescope}
{To compensate the missing flux, we use data from an imaging line survey on G28.34\,S
with the IRAM 30~m telescope at 3\,mm (see details in \citealp{feng16}).} Observations were performed in the on-the-fly mode on May 28, 2014, mapping a $1.5'\times1.5'$ area centered at $\rm 18^h42^m46^s.597$, $\rm -04^{\circ}04^{'}11^{''}.940$ (J2000).
A broad bandpass (8\,GHz  bandwidth)  covers the range of 85.8--93.6\,GHz 
with a frequency resolution of 0.195\,MHz (velocity resolution of 0.641\,$\rm km\,s^{-1}$ at 3\,mm). 
The FWHM beam of the 30~m telescope is 
 $\sim30''$  at 3\,mm. 
The $\rm1\sigma$ rms $\rm T_{mb}$ in the line free channels is 6--8\,mK at 3\,mm. 
At such sensitivity, the extent of the $\rm >5\sigma$ emission of $\rm SiO$ $J=\rm 2\rightarrow1$ is $>0.2$\,pc.
\\

We combined the large-scale $\rm SiO$ $J=\rm 2\rightarrow1$ obtained from the IRAM 30\,m with the high resolution data from NOEMA using GILDAS. The combined SiO line cube has a synthesized beam of $\rm 5.35\arcsec\times2.76\arcsec$  (PA=-174$\rm ^\circ$). The spectral resolution is smoothed to $\rm 1\,km\,s^{-1}$, and the $\rm 1\sigma$ rms value is $\rm 6.1\,mJy beam^{-1}$.

\subsection{SMA}
Observations with the Submillimeter Array (SMA\footnote{The Submillimeter Array is a joint project between the Smithsonian
Astrophysical Observatory and the Academia Sinica Institute of Astronomy and
Astrophysics, and is funded by the Smithsonian Institution and the Academia
Sinica.}, \citealt{ho04}) were carried out  in the extended (EXT) and compact (COMP) configurations at 260/270\,GHz (1.1\,mm) {on June 11 and July 19 2013, respectively}. The primary beam is 48\arcsec,  and the baseline range is 16--226 m, sensitive to structures up to $10''$.
The correlator was tuned to cover 258.1--262.0\,GHz in the lower side band and 270.0--273.9\,GHz in the upper side band, with a uniform channel width of 0.812\,MHz (0.936$\rm ~km\,s^{-1}$).
More details of the observations and data calibration are given in \citet{feng16}.

No line is detected in the image domain, although a $2.5\sigma$ signal of SiO $J=\rm 6\rightarrow5$ is seen in the $(u,v)$ spectrum. 
We made continuum images using data from 
 the combined COMP+EXT data ($\rm 2.55\arcsec\times1.51\arcsec$, PA=33$\rm ^\circ$, Figure~\ref{source}c).
The $\rm 1\sigma$ rms value is  $\rm0.97\,mJy beam^{-1}$.

\subsection{ALMA archival data}
We further obtained archival data observed with ALMA. The project number is 2011.0.00236.S, and was carried out with the compact configuration in ALMA Cycle-0. The details of the observations can be found in \cite{tan13}.  We download the calibrated data from the ALMA archive and cleaned it with natural weighting in CASA \citep{mcuullin07}. The derived angular resolution is  $\rm 2.35\arcsec\times1.98\arcsec$  ($\rm PA=-66^{\circ}$), and the maximum detectable scale is $\sim 9''$. The rms noise level is $\sim 7.4$ mJy per 0.08 km$\rm s^{-1}$ channel after continuum subtraction. Unfortunately the bandpass only covers part of the blueshifted velocity range {($\rm <69.5\,km\,s^{-1}$)} of the SiO $J=\rm 5\rightarrow4$ emission.

\section{Results}\label{sec:result}
\subsection{Spatial distribution of SiO  emission}
Figure~\ref{source}d shows the 0.06\,pc resolution, 3.4\,mm continuum image of S-A (greyscale), overlaid with the velocity-integrated $\rm SiO$ $J=\rm 2\rightarrow1$  flux intensities (contours). The 3.4 mm continuum image resolves a single dust component, which has a 2-D Gaussian deconvolved size scale of $\sim$0.08 pc ($\rm 5.04\arcsec\times2.62\arcsec$, PA=92$\rm ^\circ$).
In addition, we detected high velocity blue-shifted and red-shifted gas lobes in $\rm SiO$ $J=\rm 2\rightarrow1$, which are associated with the 3.4\,mm continuum in the W-E orientation. 
The 2-D Gaussian deconvolved size  of the blue-shifted and red-shifted gas lobes are $\rm 6.42\arcsec\times5.48\arcsec$ ($\rm PA=55^{\circ}$) and $\rm 7.76\arcsec\times5.82\arcsec$ ($\rm PA=98^{\circ}$), respectively.

The blue-shifted gas lobe seen in $\rm SiO$ $J=\rm 5\rightarrow4$ appears more compact ($\rm 2.9\arcsec\times2.2\arcsec$, PA=108$\rm ^\circ$  after 2-D Gaussian deconvolution). It is  closer to the 3.4\,mm continuum center than  $\rm SiO$ $J=\rm 2\rightarrow1$ by 2\arcsec.
This is likely due to a higher excitation condition requirement of the $\rm SiO$ $J=\rm 5\rightarrow4$ line, which better traces a localized warmer part of the blue-shifted gas. 
The ALMA data also provides a better sensitivity than the
NOEMA-30\,m observations\footnote{The ALMA $J=\rm 5\rightarrow4$
observation is $\sim$ 5.5 times more sensitive than the NOEMA-30\,m  $J=\rm
2\rightarrow1$ observations in detecting the outflow mass, when smoothing both
data into the same spatial and velocity resolution and assuming the same gas
temperature of  15\,K.}, so it can detect the highest velocity line wing of the
outflow which is not seen in the NOEMA-30\,m data.

\subsection{Position-velocity (PV) diagram}
\begin{figure*}
\centering
\begin{tabular}{cc}
\includegraphics[width=10cm]{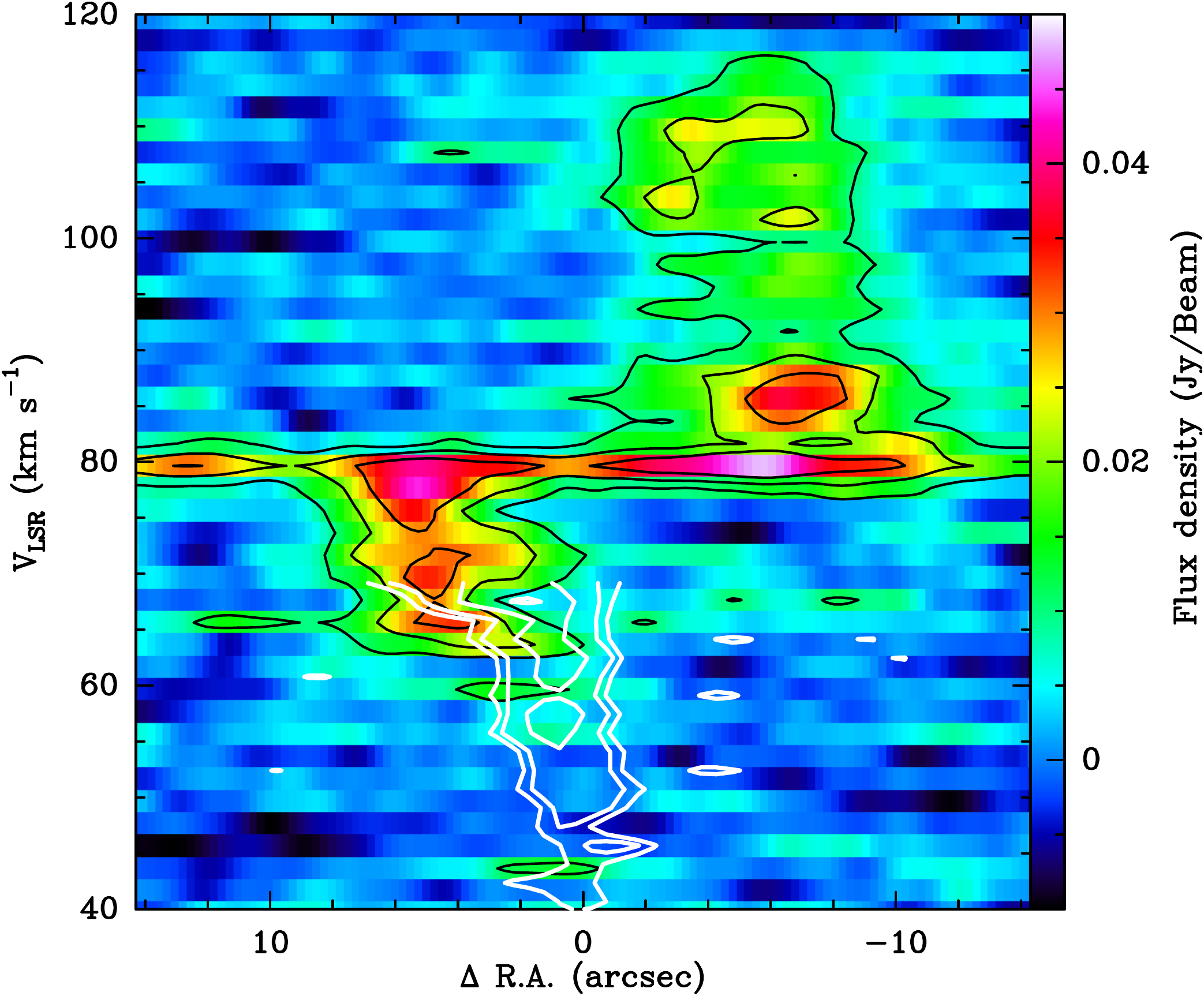}
\includegraphics[width=5cm]{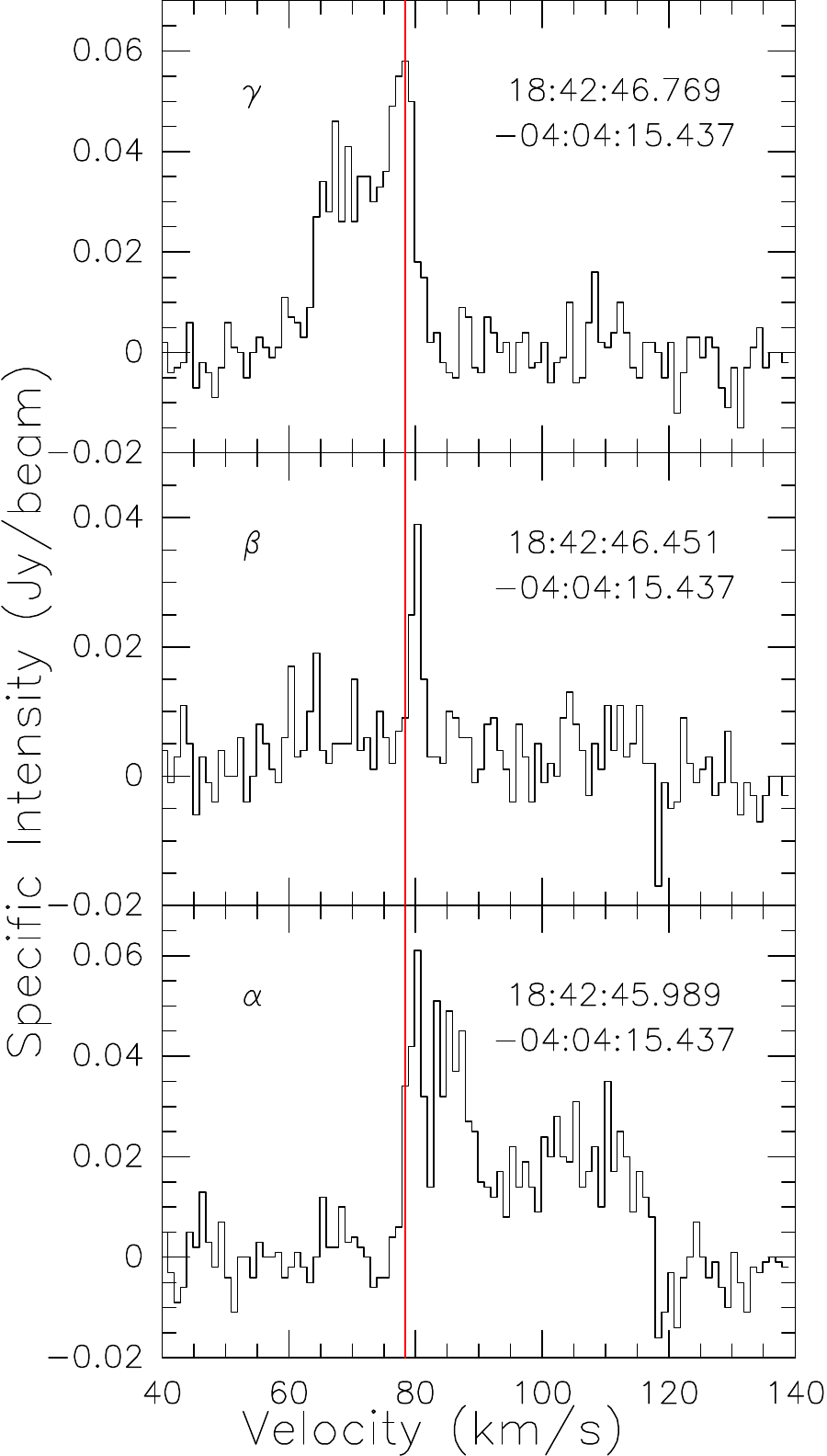}
\end{tabular}
\caption[]{Left panel: position-velocity diagram along the W-E orientation with a width of 5\arcsec.  The colormap and the black contours are extracted from the $\rm SiO$ $J=\rm 2\rightarrow1$ line, with contour levels of 0.01, 0.02, and the 0.03 $\rm Jy\,beam^{-1}$. The white contours are from the $\rm SiO$ $J=\rm 5\rightarrow4$ line ({velocity coverage up to  $\rm 69.5\,km\,s^{-1}$ by ALMA observations}), with contour levels of 0.006, 0.01, and 0.02 $\rm Jy\,beam^{-1}$. Note that the ALMA data has been convolved to the same angular resolution of the NOEMA data for comparison.  Right panel: line profiles extracted from the positions marked in Figure~\ref{source}d.
\label{pv}}
\end{figure*}

{To study the  velocity distributions of two SiO lines along the W-E extension, we convolve the  $J$=5$\rightarrow$4 data obtained  with ALMA to the same angular resolution as the  $J$=2$\rightarrow$1 data obtained with NOEMA.} 
To increase the signal to noise ratio, we take a 5\arcsec-width  slice  along the W-E orientation and extract a {PV diagram} from both  lines (Figure~\ref{pv} left panel), using the ``pvextractor" code\footnote{http://pvextractor.readthedocs.org}.
The PV diagram of both SiO transitions reveals that the higher-velocity components are located
closer to the continuum {peak} while the lower-velocity components are further
away from the continuum. This is likely because: (1) {The excitation conditions}. The
highest velocity gas near the continuum {peak} might have a higher excitation
with a higher temperature and density.  The high-$J$ SiO lines could be
enhanced by the high velocity shocks, while the low-$J$ lines are from the
entrained cocoon which has lower velocities (see similar case in HH\,211,
\citealp{hirano06,palau06,lee07});  and (2) An mechanical effect. It is also possible that the central protostar ejects outflows episodically, and
later ejecta have higher velocities; or, the outflow is still in an early stage
and the shock front decelerates because of its interaction with the ambient gas. {These may bring in spatially differentiated velocity components  from the same transition ($J=2\rightarrow1$).}

\subsection{SiO Line profiles: towards/away from the dust core}
Line profiles of SiO $J=2\rightarrow1$ vary spatially (Figure~\ref{pv} right panel). Narrow line emission (FWZI, the overall zero-intensity velocity dispersion of $\rm \sim4\,km\,s^{-1}$) with $\rm \sim +2\,km\,s^{-1}$ shifted from  the $\rm V_{lsr}$ of the cloud {is detected towards} the  continuum center. In contrast, line profiles of SiO $J=2\rightarrow1$ {towards} the emission peak of the blue-/red-shifted lobes are broader (FWZI  $\rm >30\,km\,s^{-1}$). Each profile has double intensity peaks: a more intensive peak with narrow FWHM linewidth  around the $\rm V_{lsr}$ and  a higher-velocity peak with broader line wing.
Such profiles perfectly match the prediction of  the evolutionary SiO line profiles introduced by C-shock models at the core scale \citep{jimenez09}: The low-temperature neutral fluid at the young, magnetic precursor stage produces the narrow line, and a small ion-neutral drift velocity leads to the sputtering of the grain mantles. Shock propagation and heating broadens the line profiles in the lobes. The more intensive peak at  lower-velocity is related to  the  silicon ejected from the grain, while the less intensive peak at  higher-velocity corresponds to the large accumulation of SiO in the postshock. \\

{The above observation results 
 indicate that  SiO emissions  trace the shocked gas when the W-E oriented outflow impacts on the ambient gas.}
Besides the outflow, other possible enriching mechanisms of SiO include vaporization and shattering by  grain-grain collisions  (e.g., \citealt{caselli97,guillet09,guillet11,anderl13}),  cloud-cloud collision (e.g.,   \citealt{jimenez09,jimenez10,nguyen13}).
However, previous and our observations do not provide evidences for these mechanisms. For example, neither obvious velocity gradients of SiO nor temperature gradients in $\rm NH_3$ \citep{wangy08} are detected on larger scales ($\rm >$1\,pc).  \\
\color{black}

\section{Discussion}\label{sec:discussion}
{In the following discussions,
we provide quantitative derivations of the physical properties of the unresolved high-mass molecular gas core (Section \ref{sub:core}), of the the W-E oriented SiO outflow (Section \ref{sub:outflow}), and of  the low velocity ($\rm <5\,km\,s^{-1}$) SiO emission (Section \ref{sub:2of}). 
We also discuss the physical implication of the derived quantities in Section \ref{sub:store}. 
}

\subsection{Core mass}\label{sub:core}
{
Previous studies have indicated that S-A is a cold and dense gas core.
Therefore, it is safe to assume that dust and gas are thermalized, such that dust temperature is equal to gas kinetic temperature (e.g., \citealp{juvela11}).}
  {Using the combined VLA+Effelsberg $\rm NH_3$ lines (resolution of 5\arcsec, \citealp{wangy08}) and the procedure outlined in \citet{wangk14}, we fit a gas kinetic temperature of $T_{\rm kin} \approx 15$ K for S-A.}

 We assume that the 1.1\,mm and 3.4\,mm continuum emission is dominated by thermal dust emission.  
Adopting a gas to dust mass ratio of 150 \citep{draine11}, a dust opacity of $\rm \kappa_{1.1mm}=1.0\,cm^2g^{-1}$ and $\rm \kappa_{3.4mm}=0.17\,cm^2g^{-1}$ ($\rm \kappa \propto \nu^{1.8}$, \citealp{ossenkopf94}), we estimate a core mass of $\rm \sim40\,M_{\odot}$ ($\rm \sim53\,M_{\odot}$), a $\rm H_2$ column density of $\rm \sim 5\times10^{23}\, cm^{-2}$ ($\rm \sim 6\times10^{23}\, cm^{-2}$), {and a number density  $n\rm \sim 3\times10^{6}\, cm^{-3}$ ($\rm \sim 5\times10^{6}\, cm^{-3}$)}  in the unresolved core S-A, from  the 3.4\,mm (1.1\,mm) continuum, obtained with NOEMA (SMA). 
Comparing to the 870\,$\mu$m single-dish continuum emission and assuming a spectral index in the range of 3.5--4, more than 60\% of the missing flux leads to the above core mass being a lower limit. 
{Our constraints on the gas number density imply a free-fall collapsing timescale shorter than $\rm t_{ff}=3.66\times10^7/\sqrt{n}\,yr\sim10^5$\,yr. 
Dividing core mass by the free-fall collapsing timescale, the inferred averaged accretion rate  is $\rm 10^{-3}\,M_{\odot}yr^{-1}$ or higher.}

\subsection{Physical properties of the W-E oriented outflow}\label{sub:outflow}

The {entire} G\,28.34\,S region has a low bolometric luminosity $\rm L_{bol}\sim 10\,L_{\odot}$ {(priv. communication with S. Ragan)}. 
{Assuming that this luminosity is mainly  due to accretion of a single young stellar object, 
the present mass of the central stellar object in S-A is likely in the range of $\rm 0.1\,M_{\odot}\lesssim M_*<3\,M_{\odot}$ \citep{dunham12,krumholz14}. 
The possible radius of the host (proto)star $\rm R_*$ ranges from $\rm 1.5\,R_{\odot}$ \citep{stahler80} to several tens of $\rm R_{\odot}$ (e.g., \citealp{hosokawa10,hosokawa11}).  
The present accretion rate may range 
$\rm \dot{M}_{acc}<L_{bol}R_*/(GM_*)= 10^{-7}\text{--}10^{-5}\,M_{\odot}yr^{-1}$. 

 The exact physical properties of the SiO outflows is yet uncertain, due to the unknown SiO abundance with respect to $\rm H_2$ ($\rm \chi_{SiO}$).
However, $\rm \chi_{SiO}$ may be related to the evolutionary scenario of the host powering source through the following discussion.
Our following estimations are based on two assumptions: (1) the shocked gas in the outflow is warmer than or at least has the same temperature as the core centre (15\,K); (2) $\rm SiO$ $J=\rm 2\rightarrow1$ is optically thin. 
}

 Using 2-D Gaussian fits to the gas lobes in Figure \ref{source}d, we measure the SiO $J=2\rightarrow1$ total flux of the red-shifted lobe as 4.02\,Jy and of the blue-shifted lobe as 2.06\,Jy. The H$_2$ gas mass of the double-side lobes $\rm M_{lobes}=M_{red}+M_{blue} ={2.1\times10^{-11}}/{\chi_{SiO}}\,M_{\odot}$.
 
Using the highest  outflow velocity observed from SiO $\rm J=2\rightarrow1$ ($\rm V_{max}$, $\rm 58.4\,km\,s^{-1}$ and $\rm 113.4\,km\,s^{-1}$) in each lobe and $\rm V_{lsr}$ ($\rm 78.4\,km\,s^{-1}$), the momentum of the outflow is $\rm P=M_{red}|V_{max,red}-V_{lsr}|+M_{blue}|V_{max,blue}-V_{lsr}|={6.1\times10^{-10}}/{\chi_{SiO}}\,M_{\odot}km\,s^{-1}$.
The high-velocity components are closer to the dust center, so we assume that different velocity components form different layers of gas. Comparing the  timescale $\rm t_{\it i}$ each velocity component $\rm V_{i}$ having been through to form its layer, with projection length $\rm L_{i}$ from the continuum center ($\rm t_{\it i}=L_{\it i}/|V_{\it i}-V_{lsr}|$), we  derive the maximum as the outflow dynamic timescale  $\rm t_{dyn}\sim10^4$ years, {which is shorter than the free-fall collapsing timescale of the parent dense core (Section \ref{sub:core}) by a factor of 10. This is consistent with the fact that the observed source may be so young that it is at the beginning  of star-formation.

Assuming that the outflow is momentum driven by the underlying wind, which has a speed  $\rm V_w$ on the same order as the Keplerian velocity  at stellar surface, $\rm V_k=\sqrt{GM_*/R_*}$ \citep{krumholz05}, i.e., in the range of $\rm 20\text{--}600\,km\,s^{-1}$. Taking  $\rm V_w=300\,km\,s^{-1}$, the mass-loss rate to the wind is  $\rm \dot{M}_{w}=P/(V_wt_{dyn})={4\times10^{-14}}/{\chi_{SiO}}\,M_{\odot}yr^{-1}$. We adopt a double-side mass loading ratio ($\rm \dot{M}_{w}/\dot{M}_{acc}$) as in the range of 0.1--0.3  \citep{tan14}. Therefore, possible scenarios based on the assumption of a constant mass loading are:\\
(1) A dominant outflow-powering young stellar object, which has a relatively high mass ($\rm 1\text{--}2\, M_{\odot}$), has already formed in this dense core. Assuming the typical radius of such object is  several $\rm R_\odot$, the observed low bolometric luminosity ($\rm <10\, L_{\odot}$) requires the mass accretion rate to be not very high ($\rm \dot{M}_{acc}<10^{-7}\text{--}10^{-6}\,M_{\odot}yr^{-1}$). This will imply a  low outflow mass and a high SiO abundance ($\chi_{SiO}>4\times10^{-7}$);\\
(2) If the powering young stellar object has relatively low mass ($\rm 0.1\, M_{\odot}$) and large radius (typically tens of $\rm R_\odot$), and if the accretion dominates the low bolometric luminosity ($\rm <10\, L_{\odot}$), the mass accretion rate should be high ($\rm \dot{M}_{acc}<10^{-5}\text{--}10^{-4}\,M_{\odot}yr^{-1}$). This will imply a high outflow mass and a low SiO abundance  ($\chi_{SiO}>4\times10^{-10}$).


The second scenario may be less likely. 
Assuming gas-dust temperature in the continuum peak of S-A (position $\beta$ in Figure~\ref{source}d and  Figure~\ref{pv} right panel) is 15\,K, the SiO abundance in this dense core peak is $\rm \chi_{SiO,\beta}\sim5\times10^{-12}$. 
The predominant SiO emission observed in the outflow regions over that observed from the dense core indicates a much higher SiO abundance in the outflows.
The lower limit of the SiO abundance in the outflows given by the temperature lower limit assumption\footnote{Temperature in the outflow may be enhanced ($\rm>$15\,K) because of the shocks (e.g., \citealt{lopez16}).} (15\,K), velocity-integrated SiO flux intensity, and the $3\sigma$ continuum intensity (e.g., at positions $\alpha$ and $\gamma$), is $\rm \chi_{SiO,\alpha\sim\gamma}>10^{-9}$, which is higher than the estimates given by the second scenario.

Considering the averaged accretion rate ($\rm 10^{-3}\,M_{\odot}yr^{-1}$, see Section \ref{sub:core}) and the maximum outflow dynamic timescale ($\sim10^{4}$ yrs), the assumed $\rm 1\text{--}2\, M_{\odot}$ young stellar object in the first scenario may be reasonable.
However, the derived SiO abundance from the first scenario is slightly higher than that in other observed shocked region ($\rm 10^{-10}-10^{-7}$, e.g., \citealt{mikami92,zhang95,hirano01,gusdorf08,leurini14}).
The derived low mass accretion rate is also puzzling.
The issues of too high SiO abundance and too low accretion rate, can be alleviated if we consider the following scenario:\\ 
(3) A dominant outflow-powering young stellar object with a relatively high mass ($\rm 1\text{--}2\, M_{\odot}$) has already formed and is accreting episodically\footnote{Episodic accretion may occur when the gas core is collapsing onto the forming, approximately Keplerian-rotating disk.
If the mass accretion onto the disk is much faster than the accretion rate of the host (proto)star, which is very likely to be the case during the collapsing phase, then the accumulated mass can make the disk become Toomre unstable \citep{toomre64}, which triggers gravitational instability and the episodic accretion  (e.g.,  \citealp{vorobyov13,vorobyov15,liu16}).
The episodic scenario  therefore may imply a forming high-mass star accreting via Keplerian-rotating disk (see also \citealp{johnston15}).
A immediate consequence of the potential episodic accretion, is that the bolometric luminosity may not always be a good indicator for the evolutionary stage.}. 
In such case, the region may have high luminosity and high mass accretion during the dominant outflow eruptions, while it is  back to a relatively quiescent status with temporarily low luminosity and low mass accretion rate  currently (e.g., \citealp{liu16}). 
The potentially much higher averaged accretion rate in the past (up to $\rm <10^{-3}\,M_{\odot}yr^{-1}$) can bring in a more reasonable SiO abundance.
The better estimates of SiO abundance in the future observations by observing multiple transitions of more molecular species, may help discern the scenarios (2) and (3).

Although we cannot exclude that S-A just forms a few low-mass stars, the large gas reservoir is indicative of a high-mass star-forming core.

Finally, based on the aforementioned assumptions,  if taken  $\chi_{SiO}\sim 10^{-7}$, the mass entrainment rate of the outflow is $\rm M_{lobes}/t_{dyn}\sim10^{-8}\,M_{\odot}yr^{-1}$, the kinetic energy is $\rm E_{lobes}=1/2(M_{red}|V_{max,red}-V_{lsr}|^2+M_{blue}|V_{max,blue}-V_{lsr}|^2)\sim2\times10^{42}\,ergs$, and the mechanical luminosity is $\rm E_{lobe}/t_{dyn}\sim10^{-3}\,L_{\odot}$. If taken $\chi_{SiO}\sim 10^{-9}$, the above parameters should be two magnitudes higher. However,  these values indicates the W-E outflow from S-A is less intense than the typical outflows from the more evolved high-mass star-forming regions \citep{beuther02b,arce07,zhang15}.
}

\subsection{A second outflow?}\label{sub:2of}
\begin{figure*}
\centering
\begin{tabular}{cc}
\includegraphics[width=15cm]{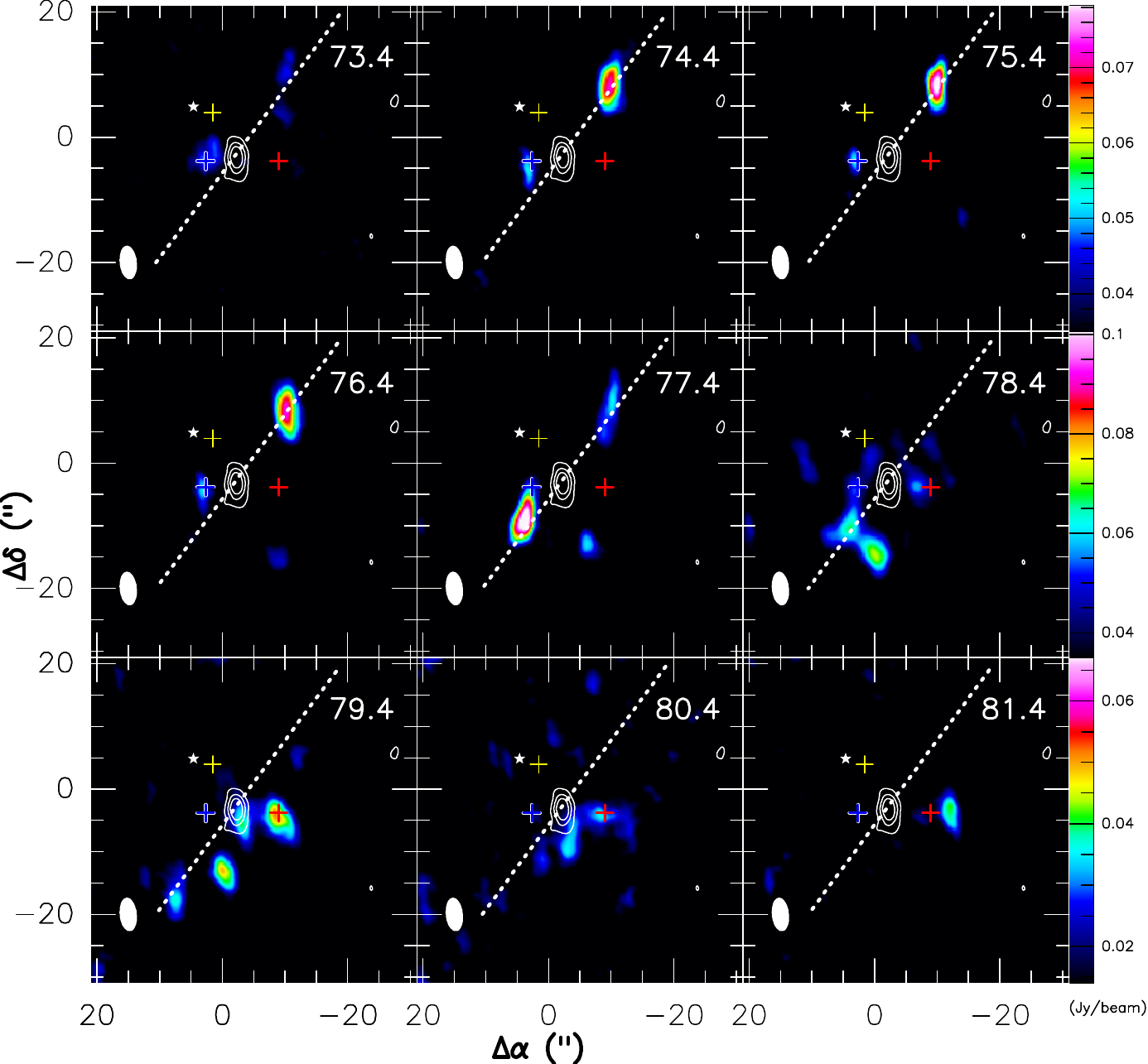}
\end{tabular}
\caption[]{Channel maps of $\rm SiO$ $J=\rm 2\rightarrow1$ (colormap) in the velocity range of $\rm V_{lsr}$$\rm _{-5}^{+3}\,km\,s^{-1}$. The white contours show the 3.4\,mm continuum emission. The blue (red) cross shows the blue-shifted (red-shifted) gas lobe. The yellow cross denotes the $\rm H_2O$ maser. The star marks the 1.2\,mm continuum source S-B. The white dashed line indicates the elongation of the NW-SE structure. \label{channel}}
\end{figure*}

Besides the W-E oriented high-velocity outflow, we note that the spatial distribution of  the low-velocity $J=\rm 2\rightarrow1$ line emission shows a NW-SE elongation, which is almost perpendicular to the major axis of the large scale filament. Channel maps in the velocity range of $\rm V_{lsr}$$\rm_{-5}^{+3}\,km\,s^{-1}$ reveal a second pair of bright blue-shifted and red-shifted gas lobes, which are centered on the unresolved 3.4 mm continuum peak and are aligned at a position angle of $\rm PA\sim-53^{\circ}$ (Figure~\ref{channel}). The channel maps further indicate that the SE lobe is coincident with the blue-shifted lobe of the W-E outflow at $\rm \sim77.4\,km\,s^{-1}$. Its origin is uncertain, and possible mechanism for this elongation can be (1) a second outflow from  the unresolved protostar(s) in S-A; (2) 
large-scale shock remnants  \citep{jimenez04}. Although there is no significant evidence to rule out the second possibility, the gas distribution from our observations favours the first explanation.

In addition, a water maser ($\rm 59.5\,km\,s^{-1}$) coincides with the ridge between S-B and the shocked gas of S-A. Since this maser is variable \citep{wangy06,wangk12}, {it is} not clear which source it relates to, {but its presence marks the existence of a protostellar outflow}.

\subsection{A global picture of the $\rm 70\,\mu$m dark G28.34\,S}\label{sub:store}
The high-mass cores S-A and S-B resolved from SMA observations {are located at} the southern edge of IRDC G28.34+0.06  \citep{feng16}.
  Aligned with the filament, they are consistent with the hierarchical fragmentation scheme found in the P1 clump \citep{wangk11,wangk12,zhang15}, which is 1\arcmin~ to the NW and  has several co-existing bipolar, jet-like, CO outflows. The W-E oriented outflow found in S-A also has other similarities to the outflows in P1. In particular, 
all outflows have orientations {almost perpendicular to} the major axis of the filament. However, the momentum, mechanical luminosity, and flow mass entrainment of the S-A outflow is {at least one magnitude} less intense than those found in P1, indicating the stellar object(s) in S-A is less evolved than the others in this IRDC. 
Despite of the different outflow properties, S-A has the similar mass and IR properties to some other IR-dark cores embedded in P1.
This indicates the stellar object(s) in S-A is less evolved than the others in this IRDC. 
Similar outflows have also been reported in other IR-dark and IR-bright cores embedded in the IRDC ``Snake'' \citep{wangk14}.




The total bolometric luminosity found in G28.34\,S is low (10\,$\rm L_{\odot}$), which makes it intriguing to find  an  outflow 
 in a 0.1\,pc-unresolved young high-mass  star-forming core such as S-A.  
More importantly, the outflows detected in P1 and S-A indicate that a {mechanical energy} feedbacks  the cores { which} are still dark at 70\,$\mu$m.  What is the mechanism that drives the outflows? What is the evolutionary status of these cores? How many protostellar objects have been formed but not yet globally heat up the parental molecular clump  {(see more discussion on localized outflow heating in G28.34-P1 in \citealt{wangk12})}?  

We note that a recent large survey on scales of 0.5--1\,pc presents detections of $\rm SiO$ $J=\rm 2\rightarrow1$ and $J=5\rightarrow4$ from many IR-quiet clumps \citep{csengeri15}.
More specifically, 25\% of those observed  IR-quiet clumps present high-velocity $J=\rm 2\rightarrow1$ line wings, indicating outflows. 
High-mass molecular gas cores may spend considerable life time in the stage which exhibits low luminosity, but already posses active star-formation and outflow activities.
The source S-A  may represent the earliest part of this evolutionary stage.

\section{Conclusion}\label{sec:conclusion}
We present NOEMA and IRAM 30\,m observations that uncover the presence of at least one bipolar outflow in a $\rm 70\,\mu$m  dark core G\,28.34\,S-A. Revealed by shocked gas as  high-velocity red-/blue shifted lobes of $\rm SiO$ $J=\rm 2\rightarrow1$, this W-E oriented outflow is also {shown by} the blue-shifted gas lobe of $\rm SiO$ $J=\rm 5\rightarrow4$ (from ALMA cycle 0 archival data). The PV-diagram of this outflow indicates {that} the higher-velocity components are closer to the continuum center. Moreover, the  momentum,  mechanism luminosity and  mass entrainment of this outflow in such a low luminosity core indicate the center source(s) {is in an early evolutionary stage of forming  high-mass protostar(s).} 
In addition, blue- and red-shifted gas reveal low-velocity elongation in the NW-SE orientation, which may be from a second outflow. {Furthermore, we discuss the possible accretion scenarios of the outflow-powering young stellar object. We argue that the molecular line emission and the molecular outflows may provide a better indication of the accretion history of the forming young stellar object and thereby the evolutionary stage, than that from a snapshot observations of the present bolometric luminosity. }

\begin{acknowledgements}
We would like to thank the SMA staff, IRAM 30\,m {staff} for
their helpful support during the performance of telescopes in service mode.
We thank Jan-Martin Winters for helping with the NOEMA data reduction.
We thank Paola Caselli, Jaime Pineda, and Sarah Ragan  for helpful discussion.  \\
This research made use of NASA's Astrophysics Data System.\\
ZYZ acknowledges support from the European Research Council (ERC) in the form of Advanced Grant, {\sc cosmicism}.
{KW acknowledges support from grant WA3628-1/1 of the German Research Foundation (DFG) through the priority program 1573 (``Physics of the Interstellar Medium'').}
\end{acknowledgements}

\bibliographystyle{apj}
\bibliography{SiO-G28-accepted.bbl}

\end{document}